\documentclass[a4paper]{jpconf}
\usepackage{graphicx}
\def\mathnew{\mathsurround=0pt}
\def\simov#1#2{\lower .5pt\vbox{\baselineskip0pt \lineskip-.5pt
       \ialign{$\mathnew#1\hfil##\hfil$\crcr#2\crcr\sim\crcr}}}

\def\siml{\mathrel{\mathpalette\simov <}}

\def\compton{\textit{Compton}\,}
\def\swift{\textit{Swift}\,}
\def\fermi{\textit{Fermi}\,}

\def\gbm{\textit{GBM}\,}
\def\lat{\textit{LAT}\,}

\def\vareps{\varepsilon}

\def\beq{\begin{equation}}
\def\enq{\end{equation}}
\def\bea{\begin{eqnarray}}
\def\ena{\end{eqnarray}}
\def\bec{\begin{center}}
\def\enc{\end{center}}

\def\blist{\begin{list}{$\bullet$}{\itemsep 0.0in \parsep 0.0in}}
\def\elist{\end{list}}
\def\bitem{\begin{list}{\arabic{enumi}.}{\usecounter{enumi} \itemsep 0.0in \parsep 0.0in}}
\def\eitem{\end{list}}

\def\s{\hbox{~s}}

\def\part{\partial}

\def\h75{h_{75}}
\def\Omh75{\Omega h^2_{75}}

\def\fun#1#2{\lower3.6pt\vbox{\baselineskip0pt\lineskip.9pt
  \ialign{$\mathsurround=0pt#1\hfil##\hfil$\crcr#2\crcr\sim\crcr}}}

\def\mnras{M.N.R.A.S.\,}

\def\apj{Astrophys.J.\,}
\def\apjl{Astrophys.J.Lett.\,}

\def\nat{Nature\,}

\def\prl{Phys. Rev. Lett.\,}

\def\prd{Phys.Rev.D\,}
\def\araa{Annu.Rev.Astron.Astrophys.\,}
\def\aap{Astron.Astrophys.\,}

\begin{document}
\title{Gamma Ray Bursts: recent results and connections to 
very high energy Cosmic Rays and Neutrinos\footnote{\footnotesize Plenary talk 
at PASCOS 12, M\'erida, Yucat\'an, Mexico, 2012; to appear in J.Phys. (Conf.Series)}}

\author{P\'eter M\'esz\'aros$^\ast$, Katsuaki Asano$^\dagger$, P\'eter Veres$^\ast$}

\address{$^\ast$ Center for Particle and Gravitational Astrophysics,
Dept. of Astronomy \& Astrophysics and Dept. of Physics,
Pennsylvania State University, University Park, PA 16802, USA}
\address{$^\dagger$ Interactive Research Center of Science, 
Tokyo Institute of Technology, 2-12-1 Ookayama, Meguro-ku, Tokyo 152-8550, Japan}

\ead{pmeszaros@astro.psu.edu, asano@phys.titech.ac.jp, puv2@psu.edu}

\begin{abstract}
Gamma-ray bursts are the most concentrated explosions in the Universe. They
have been detected electromagnetically at energies up to tens of GeV, and 
it is suspected that they could be active at least up to TeV energies. It is
also speculated that they could emit cosmic rays and neutrinos at energies 
reaching up to the $10^{18}-10^{20}$ eV range. Here we review the recent 
developments in the photon phenomenology in the light of  \swift and \fermi 
satellite observations, as well as recent IceCube upper limits on their 
neutrino luminosity. We discuss some of the theoretical models developed to 
explain these observations and their possible contribution to a very high 
energy cosmic ray and neutrino background.  
\end{abstract}

\section{Introduction}
\label{sec:intro}

Gamma-ray bursts (GRB) are detected at an average rate of a one  per day,
lasting in gamma-rays from fractions of a second to tens of minutes.
We know that these objects are distributed almost isotropically throughout 
the Universe, out to the largest cosmological distances yet sampled.  Yet,
while they are on, they far outshine all other sources of gamma-rays in the
sky, including the Sun. They are, in fact, the most concentrated and brightest
electromagnetic explosions in the Universe, the GRB prompt electromagnetic 
energy output during tens of seconds being comparable to that of the Sun over 
$\sim {\rm few} \times 10^{10}$ years, or to that of the entire Milky Way over 
a few years. Their initial $\gamma-$ ray emission is followed by an X-ray 
and an optical afterglow lasting for weeks, which during the first day can 
outshine the brightest quasars and active galactic nuclei in the Universe.

It is thought that GRB result as a consequence of a cataclysmic ``end game" in
the life of very evolved stars, where about a solar rest mass worth of gravitational 
energy is released in a matter of seconds or less within a a small region of the 
order of tens of kilometers. This is likely to be the result of the collapse of 
the core of a massive star, or in some cases from the merger of two compact 
stellar remnants, either of these scenarios ultimately leading to a stellar
mass black hole. Only a small fraction of this energy  needs to be converted into 
electromagnetic radiation to satisfy the observations. It is thought that this 
conversion occurs through the dissipation of the kinetic energy of a collimated 
relativistic jet outflow, a ``fireball", whose bulk Lorentz factors are in the 
range of $ \Gamma \sim 10^2-10^3$, expanding from the central engine which is
powered by the gravitational accretion of surrounding matter into the collapsed 
core.

The generic scenario for the production of the observed non-thermal photons typically
invokes synchrotron radiation and/or inverse Compton (IC) scattering by relativistic
electrons which have been accelerated to a power-law distribution in the shocks 
expected in the optically thin regions of  the outflow. These may be internal shocks, 
resulting in prompt $\gamma$-ray emission, and also external shocks at the termination 
of the relativistic outflow, which can explain many of the properties of the afterglows.  
Other mechanisms considered for the prompt emission are, e.g., magnetic dissipation or 
reconnection in the outflow, jitter radiation in shocks, or dissipative effects
in the photosphere where the outflow transitions to optical thinness. 

In the past few years, the \lat instrument on the \fermi spacecraft has shown that 
a substantial fraction of GRBs have photon spectra which extend at least to tens 
of GeV \cite{Fermi+12grbobs,Peer11fermirev}.  These could be due either to leptonic 
mechanisms (such as the electron synchrotron or inverse Compton mentioned above),
or they might be due to hadronic cascades. Various uncertainties hamper the analysis 
and modeling of these systems, including our the lack of knowledge about two important 
parameters of the outflow.  These are the baryon load of the outflow, and the 
magnetic ratio $\sigma$ between magnetic stresses and kinetic energy, which affect 
not only the bulk dynamics but also the mechanisms responsible for accelerating 
electrons and protons in the shocks or the dissipation region. It is the accelerated 
protons which could lead, in principle, to GRBs being luminous in cosmic rays and 
neutrinos. Under optimistic scenarios, they could be even more luminous in these
particle channels than in the commonly observed MeV electromagnetic channels.

\section{Electromagnetic phenomenology: MeV and sub-MeV}
\label{sec:em}

The prompt photon emission of GRBs, as documented already in the 1990s by the 
\compton (CGRO) satellite,  shows often rapidly variable $\gamma$-ray light-curves, 
leading to a classification into ``long" GRBs (LGRBs) whose $\gamma$-ray light curve 
lasts $2\s \siml t_\gamma \siml 10^3~\s$, and ``short" GRBs (SGRBs) for which
$t_\gamma\siml 2~\s$, although the latter can be longer at softer energies.
The first X-ray afterglows, lasting weeks or more, were discovered by the
Italian-Dutch {\it Beppo-SAX}  satellite in 1997, being acquired typically 8 hours
after the initial $\gamma$-ray trigger. The NASA \swift satellite, launched in 2004, 
was designed to study afterglows within a minute after the trigger.  

The \swift satellite has three instruments, covering the soft gamma-ray, the X-ray
and the ultraviolet/optical ranges. The gamma-ray detector locates bursts to $\sim$ 
2 arcminute accuracy and the position is used to repoint the onboard X-ray and UV/O
instruments, as well as being rapidly relayed to Earth so ground telescopes can follow 
the afterglows.  Measurements of the redshift distance and studies of host galaxies 
are generally done with large ground-based telescopes which receive the alerts 
from the spacecraft. The average \swift burst detection rate is $\sim$ 90 per year, 
of which approximately $\sim 90\%$ have detected X-ray afterglows, mostly among
the longer GRBs.  The shorter bursts, however, are harder to detect in X-rays, 
since often they fade rapidly below the X-ray sensitivity limit.  With help from
ground-based optical observations, about $\sim 60\%$ of \swift GRBs yield also an
optical afterglow detection.  One of the interesting findings of \swift was that 
the afterglow lightcurve has a complex structure. Often a fast decay is seen in 
the first 1000 s, followed by a shallow decay and then a re-steepening.

The long GRBs are found in the brightest regions of galaxies where intense star 
formation occurs.  Those which occur near enough are generally found in association 
with a (simultaneous) supernova of Type Ib or Ic (in more distant LGRB, the 
supernova is expected to be too faint to be detected). These facts support the 
view that LGRB are caused by the central core of a massive star collapsing to a 
compact object such as a black hole, or possibly a magnetar.
LGRBs are extremely bright in both their gamma-ray prompt emission and their 
multiwavelength afterglow.  This makes them unique tools for studying the 
high-redshift universe. For instance GRB 090423, at a redshift $z=8.2$ is the source 
with the largest spectroscopically determined redshift. Such high redshift bursts 
provide information about the universe at a time when it was only a few percent of 
its current age, providing information about the process of re-ionization of the
intergalactic medium and the chemical evolution of the universe.  LGRBs also 
contribute to determining the star formation history history of the Universe, since 
they are the endpoints of the lives of massive stars and their rate is approximately 
proportional to the star formation rate.  

Previous to {\swift}'s launch the origin of short GRBs was very poorly constrained, 
since no afterglows had been detected to localize them. This changed in 2005 when 
\swift and {\it HETE-2} detected afterglows leading to a precise localization for 
several SGRBs, 
the total number by now being significant.
Unlike long bursts, the evidence indicates that SGRBs originate in host galaxies 
with both low and high star formation rates, i.e. old and young stellar populations.
These host properties are substantially different than those of long bursts, 
indicating a different origin,
and, furthermore, nearby SGRBs show no evidence for simultaneous supernovae, 
both properties being very different than for long bursts.  These results support 
the interpretation that SGRBs arise from an old population of stars and are 
probably due to mergers of compact binaries, such as double neutron stars or 
neutron star - black hole binaries. 
\swift observations have also revealed, in about 25\% of SGRBs, long ($\sim 100\s$) 
lightcurve tails with softer spectra than the first short prompt emission episode,
which has a harder spectrum.
The localization by \swift of short GRBs, if they are indeed compact binary  mergers, 
could also help narrow the search window for gravitational waves from such binaries,
which would lead to a great scientific payoff for gravitational physics, as well as
for studies of progenitor stellar types and neutron star equations of state.

\section{Electromagnetic phenomenology: GeV and above}
\label{sec:gevobs}

\begin{figure}[htb]
\begin{minipage}{0.8\textwidth}
\includegraphics[width=1.0\textwidth,height=3.0in,angle=0.0]{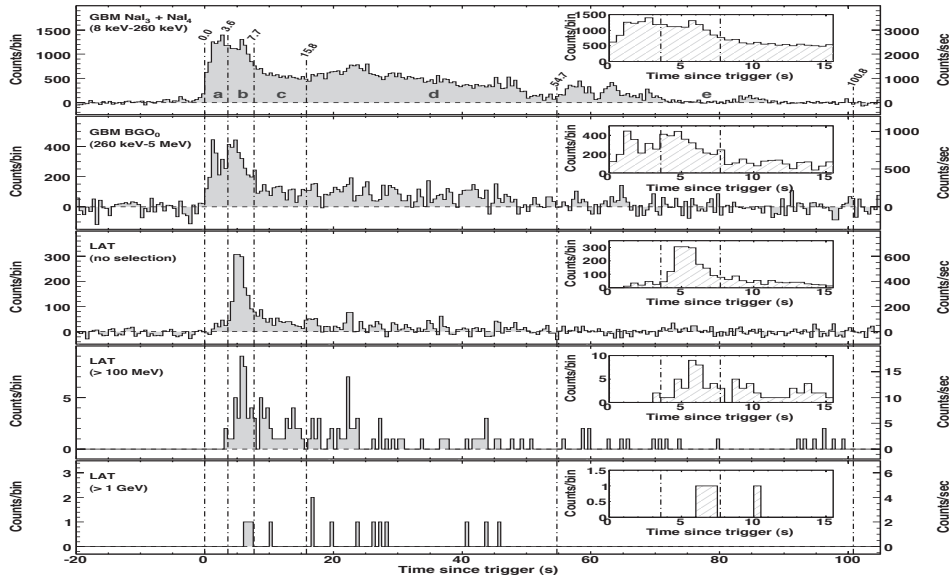}
\end{minipage}
\hspace{5mm}
\begin{minipage}[t]{0.15\textwidth}
\vspace*{-1.5in}
\caption{Light curves of GRB080916C with the \gbm (top two curves) and \lat
(bottom three curves)\cite{Abdo+09-080916}.}.
\end{minipage}
\label{fig:080916-lc}
\end{figure}

The \fermi spacecraft, launched in late 2008, started detecting GRBs with two
instruments, the Large Area Telescope (\lat, 20~MeV to $>300$~GeV) and the 
Gamma-ray Burst Monitor (\gbm, 8~keV to 40~MeV), which jointly cover  more than 
seven decades in energy. The low energy \gbm triggers on bursts at a rate of 
about 250 yr$^{-1}$, of which $\sim 80\%$ are LGRBs and $\sim 20\%$ are  SGRBs, while 
the high energy instrument, \lat, detects bursts at a rate of  $\sim 10$ yr$^{-1}$.
Of the latter, the \lat detects about twice as many at energies $\ge 100$ MeV than
it does at energies $\ge 1$ GeV.
An interesting and unexpected behavior is that in many cases 
the GeV emission starts with a noticeable delay after the MeV emission. 
E.g. in  GRB 080916C, the GeV emission appears only in a second pulse, 
delayed by $\sim 4$ s relative to the first pulse (visible only in MeV);
\cite{Abdo+09-080916}, see Fig. \ref{fig:080916-lc}.
Such a delay is present also in short bursts, such as GRB 090510
\cite{Abdo+09-090510,Ackermann+10-090510},
where it is a fraction of  a second. Such a soft-to-hard spectral evolution is 
clearly seen in the brightest \lat bursts, and also to various degrees 
in most other weaker \lat bursts. 

Perhaps the most exciting, or exotic, consequence of the observed delays
between the \lat GeV emission and the \gbm MeV emission is that it can
be used to set robust constraints on effective field theory formulations
of quantum gravity. In particular it rules out a first order dependence on
$(E_\gamma/E_{\rm Planck})$ of any Lorentz invariance violating (LIV) terms,
using GRB~090510 data~\cite{Abdo+09-090510}. This result is robust, the
limits getting even more stringent if there additional astrophysics causes
for the delay, which indeed are expected. 

\begin{figure}[htb]
\begin{minipage}{0.75\textwidth}
\includegraphics[width=1.0\textwidth,height=3.0in,angle=0.0]{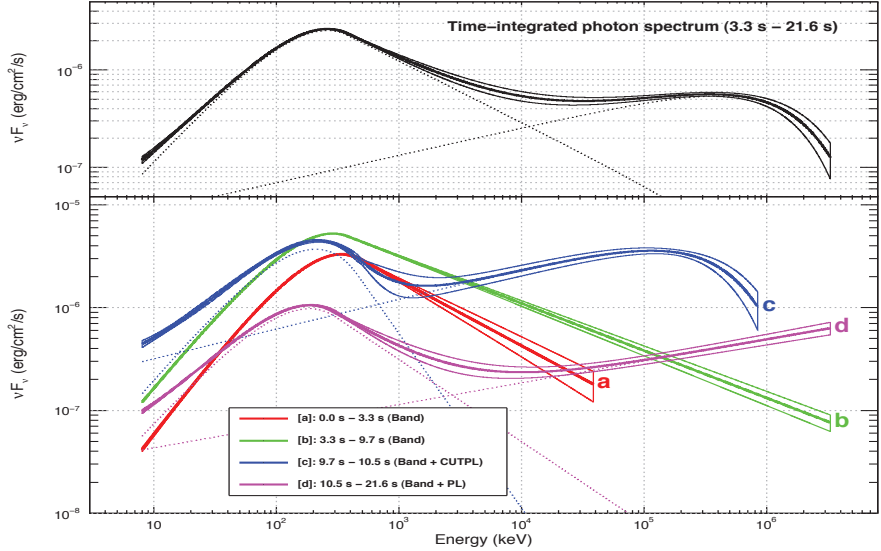}
\end{minipage}
\hspace{5mm}
\begin{minipage}[t]{0.2\textwidth}
\vspace*{-1.5in}
\caption{Spectra of GRB090926A from \fermi at four different time intervals, 
a= [0.0-3.3s], b= [3.3-9.7s], c= [9.7-10.5s], d= [10.5-21.6s] 
\cite{Ackermann+11-090926}.}
\end{minipage}
\label{fig:090926A-spec}
\end{figure}

In some burst, such as GRB 080916C, the broad-band gamma-ray spectra appear, to
within statistical accuracy, as simple ``Band" type (broken power law) functions 
in {\it all} time bins (similar to the spectrum in time bin [a] of Fig. 
\ref{fig:090926A-spec}). The absence of statistically significant evidence for a 
distinct second high energy spectral component in this and some other \lat bursts 
was initially puzzling, since naively such an extra component is expected from 
inverse Compton up-scattering or from hadronic cascades. 
However, subsequent observations, e.g. of GRB090510~\cite{Abdo+09-090510,
Ackermann+10-090510} and GRB~090902B~\cite{Abdo+09-090902}, have shown a second 
hard spectral component, extending above 10~GeV, in addition to the common Band 
spectral component dominant in the 8 keV-10 MeV band. In some cases, such as
GRB~090926A (Fig. \ref{fig:090926A-spec}), this second hard component also shows a 
downturn around a few GeV. 
%
%


An exciting discovery by the \fermi \lat was the detection of 
GeV emission from two short bursts (GRB~081024B~\cite{Abdo+10-081024} 
and GRB~090510~ \cite{Ackermann+10-090510}), whose general behavior 
(including a GeV delay) is qualitatively similar to that of long bursts.
The ratio of short to long GRBs is $\sim 10-20$\%, 
and while the statistics on short GRBs are still small, it appears that the ratio 
of the \lat energy fluence to the \gbm fluence is $> 100\%$ for the short bursts as 
compared to $\sim 5$ -- 60\% for the long bursts. Thus, although fewer in number, future 
large ground-based Cherenkov such as {\it CTA} \cite{CTA} and {\it HAWC} \cite{HAWC}
 may be able to detect short bursts.  

A remarkable feature of both long and short GRBs is that the $\ge 100$~MeV
emission generally lasts much longer than the $\siml 1$~MeV emission.
The flux of the long-lived \lat emission decays as a power law with time, which is
more reminiscent of the smooth temporal decay of the afterglow X-ray and optical 
fluxes, rather than the variable temporal structure in the prompt keV--MeV flux. 

Interestingly, the \lat detects only $\siml 10\%$ of the bursts triggered 
by the \gbm which are in the common \gbm-\lat field of view.
This may be related to the fact that the {\it LAT}-detected GRBs, both long and 
short, are generally among the highest fluence bursts, as well as being
among the intrinsically most energetic GRBs. For instance, GRB~080916C 
was at $z=4.35$ and had an isotropic-equivalent energy of $E_{\gamma, \rm iso}
\approx 8.8\times 10^{54}$~ergs in $\gamma$ rays,  the largest ever measured
from any burst~\cite{Abdo+09-080916}. The long \lat bursts GRB~090902B
~\cite{Abdo+09-090902} at $z=1.82$ had $E_{\gamma, \rm iso}\approx 
3.6\times 10^{54}$~ergs, while GRB~090926A~\cite{Ackermann+11-090926}
at $z=2.10$ had $E_{\gamma, \rm iso}\approx 2.24\times 10^{54}$~ergs.
Even the short burst GRB~090510 at $z=0.903$ produced, within the first 2~s, an 
$E_{\gamma, \rm iso}\approx 1.1\times 10^{53}$~ergs~\cite{Ackermann+10-090510}.

\section{Leptonic GRB Models}
\label{sec:lep}

The standard scenario of the mechanics of a burst is that the rotating debris 
falling into the central black hole leads, via a shear or turbulent dynamo
mechanism, to extremely strong magnetic fields, which couple the debris to the 
rotating black hole and, like super-strong rubber bands, extract the rotational 
energy of the black hole and pump it into a jet, which becomes highly relativistic 
and collimated into a 5-10 degree angular extent (a similar jet is possible if the
central engine is a highly magnetized neutron star).

The energy of the jet is initially mainly in the kinetic energy of its motion, 
and as it moves away from the black hole, the initially large particle density 
in it decreases until at a ``photospheric" radius the photon mean free path becomes 
larger than the jet dimension and the photons trapped in the jet can escape freely.  
However, if the jet energy is still mainly bulk kinetic energy at the photosphere, 
this escaping radiation would be weak, unless a substantial fraction of the bulk
kinetic energy has been dissipated into random energy of charged particles and 
radiated. This can be achieved if the kinetic energy is dissipated beyond the 
photosphere in shocks, either internal shocks within the jet itself \cite{Rees+94is}, 
or external shocks \cite{Rees+92fball}, as the jet is decelerated by external matter.  
Charged electrons bouncing across these shocks can then be accelerated via the Fermi 
mechanism to a relativistic power law energy distribution, and can produce a 
non-thermal photon spectrum via synchrotron or inverse Compton radiation, which 
approximates the observed Band type broken power law spectra. This is the ``standard
shock leptonic" model. The simple internal shock interpretation of the prompt MeV 
emission, however, has typically a low radiative efficiency, while the observed 
spectra sometimes disagree with a straightforward synchrotron interpretation, which 
has motivated searches for alternative interpretations of the origin of the
prompt MeV emission (see below).

The external shock is generally expected to be accompanied by a reverse shock, which 
can produce a prompt optical emission \cite{Meszaros+97ag} at the time the jet 
deceleration begins. This has been detected in a number of bursts, with robotic 
ground-based telescopes such as ROTSE and others \cite{Gehrels+09araa}.  As the jet 
keeps decelerating by virtue of sweeping up increasing amounts of external matter, 
the bulk Lorentz factor of the shock decreases as a power law of the distance,
and the resulting synchrotron radiation becomes a long lasting, fading X-ray, optical 
and radio afterglow \cite{Meszaros+97ag}, whose predicted detection allowed the first 
measurements of host galaxies and redshift distances \cite{Gehrels+09araa}. 
The external shock interpretation of the late afterglow is quite robust overall. 
However, there is debate about some of the more detailed features seen in the first 
few hours of the afterglow by \swift, such as X-ray steep decays followed by a flat 
plateau and occasional large flares, and various proposed extensions of the basic
external shock picture continue to be tested.

The above internal and external standard leptonic shock model is generally used 
also for interpreting the \fermi data on individual \fermi \lat bursts, e.g. 
\cite{DePasquale+09-090510ag,Corsi+10-090510}, etc.  Broader formulations 
of the shock leptonic scenario attempting to cover \lat bursts in general were
discussed by \cite{Ghisellini+10grbrad}, where the GeV emission arises from
a fast cooling forward shock, and by \cite{Kumar+09gevfs,Kumar+10fsb}, where
the forwards shock is assumed to be slow cooling (referring to the ratio
of radiative cooling time to jet dynamic time). In such models, it is argued 
that the external shock GeV emission would naturally start with a delay relative
to the prompter MeV emission, assuming the latter arise from internal shocks
(see also \cite{Meszaros+94gev}).

\begin{figure}[htb]
{
\begin{minipage}{0.7\textwidth}
\includegraphics[width=1.0\textwidth,height=3.0in,angle=0.0]{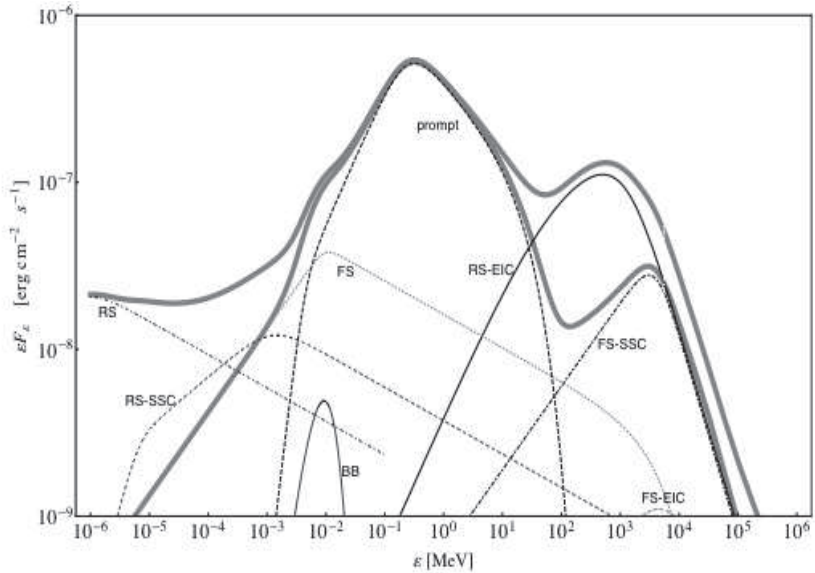}
\end{minipage}
\hspace{3mm}
\begin{minipage}[t]{0.25\textwidth}
\vspace*{-1.4in}
\caption{A magnetically dominated leptonic model, with parameters typical of
\fermi LAT GRBs.  The MeV Band spectrum is due to photospheric emission, there
are no internal shocks, and the external reverse and forward shock upscatter
the MeV  spectrum into the GeV range \cite{Veres+12mag}.}
\end{minipage}
\label{fig-ver12-two-one}
}
\end{figure}
However, taking into account the constraints provided by the Swift MeV and X-ray 
observations, it is clear that at least during the prompt emission, there must 
be an interplay between the shorter lasting mechanism providing the MeV radiation 
and the mechanism, or emission region, responsible for the bulk of the longer 
lasting GeV radiation \cite{Toma+11phot,Veres+12mag}.  The interplay between the 
two regions involves a number of subtleties, and taking into account the spatial
structure by means of multi-zone models, the inverse Compton scattering by an outer
shock of the MeV radiation from an assumed inner source of Band-like MeV photons can 
give raise to the right delays \cite{Asano+11grbtemp}. These studies avoid specifying 
a model of the prompt emission origin. If one assumes that this is due to an internal 
shock, that is open to critique because of its radiative inefficiency and sometimes 
inconsistent predicted spectra. This problem can be resolved if the prompt MeV Band 
spectrum is due to an efficient  dissipative photosphere with an internal shock 
upscattering the MeV photons at a lower efficiency, giving the delayed GeV spectrum 
\cite{Toma+11phot}. Alternatively, for a magnetically dominated outflow, where internal 
shocks are not expected, an efficient dissipative photospheric Band spectrum can be 
up-scattered by the external shock and produce the observed delayed GeV spectrum 
\cite{Veres+12mag} (see Fig. \ref{fig-ver12-two-one}).

\section{Hadronic GRB Models}
\label{sec:had}

\begin{figure}[h]
\begin{minipage}{18pc}
\includegraphics[width=18pc]{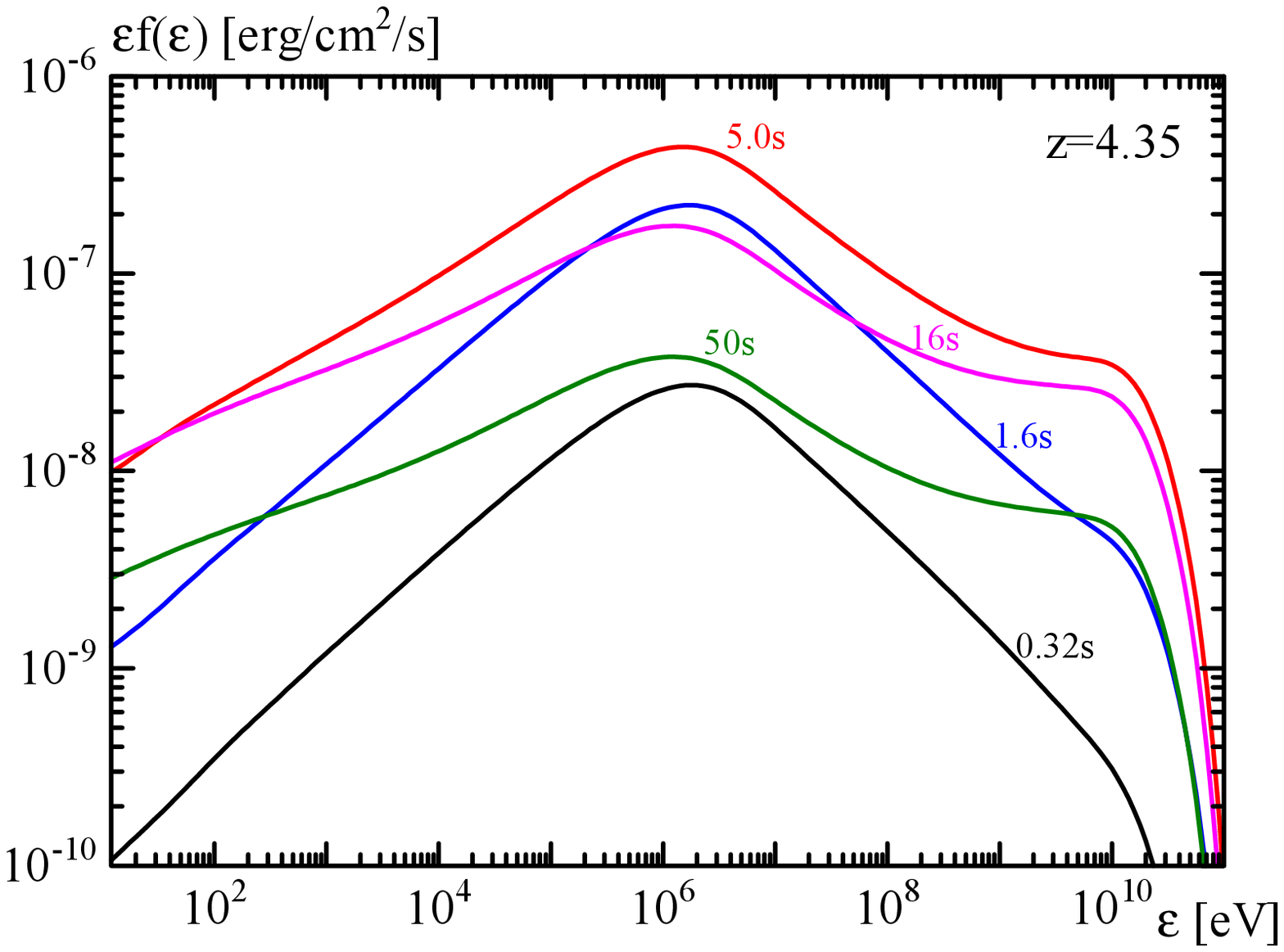}
\caption{\label{fig-asa12f11}
Hadronic GRB model Monte Carlo simulations: time evolution of the observable 
photon spectral radiation from hadronic cascades for typical \fermi \lat parameters, 
electron synchrotron producing a Band MeV spectrum and hadronic cascade secondaries 
producing the GeV spectrum as well as a low energy component \cite{Asano+12grbhad}.}
\end{minipage}\hspace{1pc}%
\begin{minipage}{17pc}
\includegraphics[width=17pc]{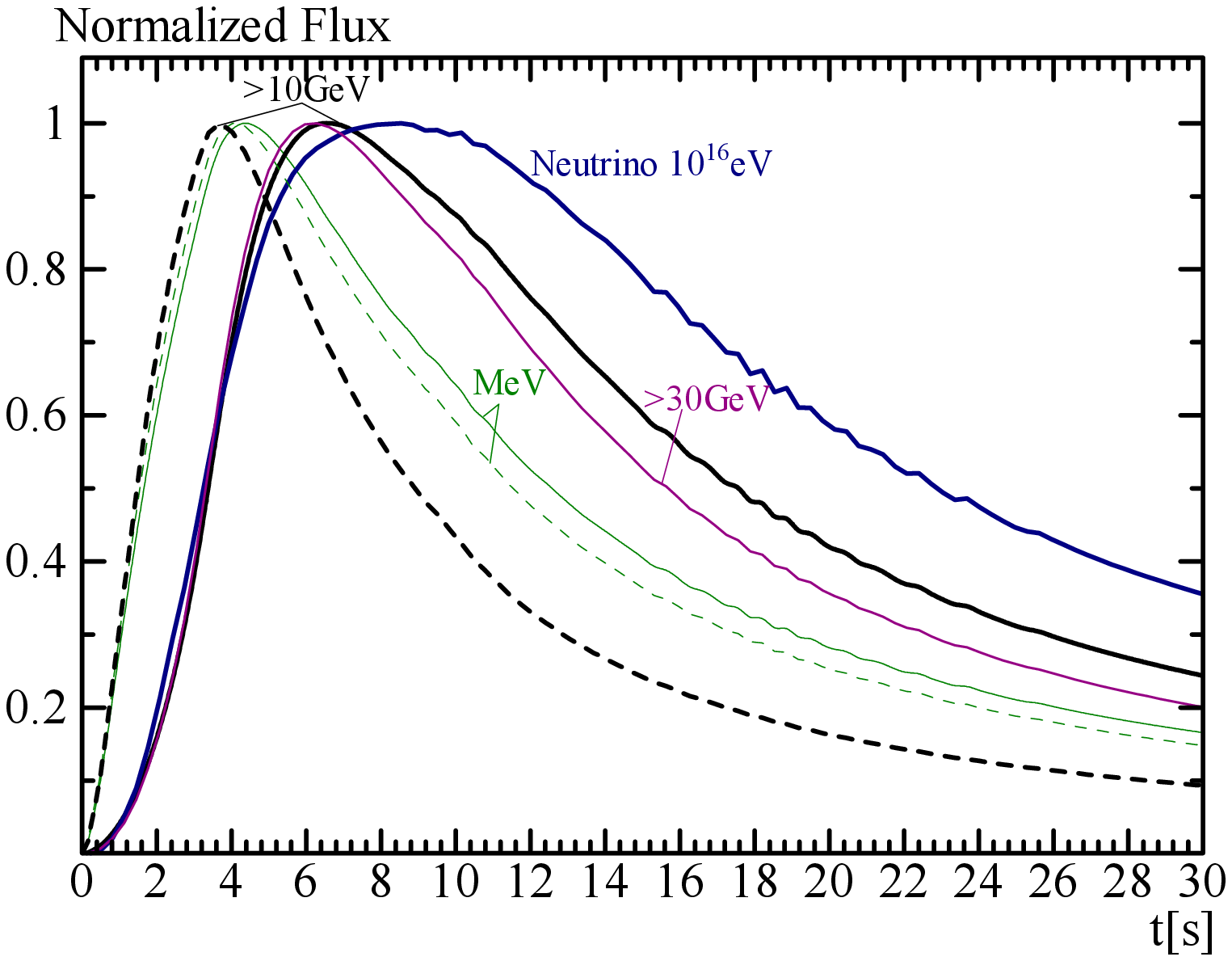}
\caption{\label{fig-asa12f13} Monte Carlo photon and neutrino light curves
of the same hadronic model (full lines) and a similar leptonic model (dashed lines)
for a bright \fermi-\lat burst at $z=4.35$, showing the expected delay between the MeV
and GeV photon lightcurves and that of the neutrinos \cite{Asano+12grbhad}.}
\end{minipage}
\end{figure}

If GRB jets are baryon loaded, the charged baryons are likely to be co-accelerated
in shocks, reconnection zones, etc., and hadronic processes would lead to both
secondary high energy photons and neutrinos. Monte Carlo codes have been developed  
to model hadronic effects in relativistic flows, including $p,\gamma$ cascades, 
Bethe-Heitler interactions, etc.  E.g., one such code \cite{Asano+09grb,Asano+09-090510} 
was used to calculate the photon spectra in GRBs from secondary leptons resulting from
hadronic interactions following the acceleration of protons in the same shocks
that accelerate primary electrons. The code uses an escape probability formulation
to compute the emerging spectra in a steady state, and provides a detailed
quantification of the signatures of hadronic interactions, which can be compared
to those arising from purely leptonic acceleration. Spectral fits of the \fermi \lat
observations of the short GRB 090510 were modeled by \cite{Asano+09-090510} as electron 
synchrotron for the MeV component and photohadronic cascade radiation for the 
GeV distinct power law component. More generally, calculations used an advanced version
of the above code show that hadronic models can describe GRB spectra where a second, 
harder photon spectral component arrives later (Fig. \ref{fig-asa12f11} 
because of the time delay needed for hadrons to be accelerated to high energies
and for cascades to develop. A prediction of such calculations is that the $\nu_\mu$
spectrum is harder than that in the models assumed in the recent IceCube papers 
\cite{Abbasi+11-ic40diff,Abbasi+12grbnu-nat}, and satisfies the constraints posed
by those papers. The neutrino light curve expected from the charged pion decays 
also shows a delay relative to the MeV photon light curve 
(Fig. \ref{fig-asa12f13}).

Hadronic interactions can also have interesting implications for GRB prompt optical 
flashes, observed in some bursts.  As discussed in \cite{Asano+10optex}, besides 
the usual Band MeV spectrum produced by leptonic mechanisms, the acceleration
of hadrons leads to secondaries whose radiation results not only in a GeV component 
but also to prompt synchrotron radiation in the optical band.  This could, in principle,  
explain the observed ``naked eye" 5th magnitude optical flash of GRB 080319B 
discussed in \cite{Racusin+08nakedeye}.

In addition to photo-hadronic interactions, also hadronic binary collisions may be 
important, both for an efficient bulk kinetic energy dissipation and for shaping 
the photon spectrum.  The baryons in a jet will be mainly protons ($p$) and neutrons 
($n$), especially if heavy elements are photo-dissociated. The protons are coupled to 
the radiation during the acceleration phase but the neutrons are carried along only
thanks to nuclear ($p,n$) elastic collisions, whose characteristic timescale at some 
point becomes longer than the expansion time. At this point the $p$ and $n$ relative
drift velocity $v$ approaches $c$, leading to the collisions becoming inelastic, $p+n 
\to \pi^+,\pi^0$, in turn leading to positrons, gamma-rays and neutrinos 
\cite{Bahcall+00pn}. Such inelastic $(p,n)$ collisions can also arise in jets where 
the bulk Lorentz factor is transversely inhomogeneous \cite{Meszaros+00gevnu}, 
e.g. going from large to small as the angle increases, as expected intuitively 
from a jet experiencing friction against the surrounding stellar envelope. 
In such cases, the neutrons from the slower, outer jet regions can diffuse
into the faster inner regions, leading to inelastic $(p,n)$ and $(n,n)$  collisions
resulting again in pions. An interesting consequence of either radial or tangential
$(n,p)$ drifts is  that the decoupling generally occurs below the scattering photosphere, 
and the resulting positrons and gamma-rays deposit a significant fraction of the 
relative kinetic energy into the flow, reheating it \cite{Beloborodov10pn}. 
Internal dissipation below the photosphere has been advocated, e.g.  \cite{Rees+05phot} 
to explain the MeV peaks as quasi-thermal photospheric peaks \cite{Ryde+10phot090902,
Peer+11bb}, while having a large radiative efficiency. Such internal dissipation is 
naturally provided by $(p,n)$ decoupling, and numerical simulations 
\cite{Beloborodov10pn} indicate that a Band spectrum and a high
efficiency is indeed obtained, which remains the case even when the flow
is magnetized up to $\vareps_B =2$ \cite{Vurm+11phot}, while keeping the dynamics
dominated by the baryons. These numerical results were obtained for nominal cases
based on a specific radial $(n,p)$ velocity difference, although the phenomenon
is generic.

\section{Cosmic rays from GRB}
\label{sec:cr}

A GRB origin of extragalactic ultra-high energy cosmic rays (UHECR)
at energies $10^{18.5}-10^{20.5}$ eV is in principle possible, 
if the lower  energy cosmic rays are provided by other, e.g.
galactic sources, such as supernova remnants \cite{Waxman95cr,Vietri95cr}. 
This is because the The maximum energy for a particle of shock accelerated
particle of charge $Z$ is $E \le \beta ZeBR$, which is $\sim 10^{20}$ eV for 
a proton in a typical GRB shock. However, only the highest energy range can 
be supplied by GRB, mainly because the spectrum is expected to be
$\propto E^{-2}$. This, combined with the low energy galactic component 
would yield approximately the observed  $E^{-2.7}$ in the sub-GZK range. 
A steeper production spectrum in GRB, such as $E^{-2.7}$ or even $E^{-2.3}$, 
in amounts enough to explain the UHECR observations in the $10^{15}-10^{18}$ 
eV range would be too energy demanding for a stellar collapse or merger 
GRB model, under usual intergalactic propagation conditions 
(c.f.  \cite{Wick+04grbcr}).

In the most common version of the GRB scenario the UHECR
are considered to be protons accelerated in GRB internal shocks
\cite{Waxman95cr,Waxman+04cr}, while another version attributes them
to external shocks \cite{Vietri95cr,Dermer02crnu}. An important caveat
of the internal shock UHECR scenario is that it assumes that the GRB prompt 
gamma-ray emission is due to such internal shocks. Although this is the 
leading work-horse scenario, there is no strong proof so far for this
(as there is for external shocks). In fact, there are problems with the 
radiative efficiency and the electrons synchrotron spectrum of internal
shocks (see previous sections), which put internal shocks into doubt, at 
least in their simple form usually assumed for UHECR GRB models.
(Such problems may be solved in alternative internal shocks or slow
magnetic dissipation models \cite{Murase+12reac}).

A relevant development is the evidence accumulating from the Pierre Auger
Observatory on the depth of shower penetration, as well as on the fluctuations
of this quantity and the muon content of the showers, which suggests that in 
the range $10^{18}-10^{20}$ eV the UHECR chemical composition acquires an 
increasing contribution from heavy nuclei \cite{Auger+10-compPRL,Auger+10-compPRL}.
In a baryonic GRB jet, where magnetic fields are dynamically sub-dominant, the 
pressure is provided by radiation, and this is expected to photo-dissociate any 
heavy elements down to $p,~n$ and $He$ \cite{Lemoine02grbnuc,Beloborodov03grbnuc}.  
However, if magnetic fields are dynamically dominant, they would provide the 
dominant pressure, the internal radiation field being lower, and in such GRB jets
nuclei can survive \cite{Metzger+11magcr,Horiuchi+12nucjet}. If GRB are the 
sources of UHECR, this is another argument suggesting that they are magnetically
dominated.

\section{High energy neutrinos from GRB}
\label{sec:nu}

If protons are accelerated in GRB shocks, these would interact with 
the observed photons mainly near the $\sim$MeV peak energy in the 
GRB spectrum, chiefly through the $\Delta^+$ resonance, $p\gamma \rightarrow \Delta^+$.
The threshold condition to produce a $\Delta^+$ is $E_p E_{\gamma}
= 0.2 \Gamma_i^2$ GeV$^2$ in the observer frame, which corresponds
to a proton energy of $E_p = 1.8 \times 10^{7} E_{\gamma, {\rm
MeV}}^{-1} \Gamma_{300}^{2}$ GeV. The short-lived $\Delta^+$
decays either to $p\pi^0$ or to $n \pi^+ \rightarrow n \mu^+
\nu_{\mu} \rightarrow n e^+ \nu_e {\bar \nu}_{\mu} \nu_{\mu}$ with
roughly equal probability. It is the latter process that produces
high energy neutrinos in the GRB fireball, contemporaneous with
the $\gamma$-rays \cite{Waxman+97grbnu}. The secondary $\pi^+$ receive $\sim
20\%$ of the proton energy in such an $p\gamma$ interaction and
each secondary lepton roughly shares 1/4 of the pion energy. Hence
each flavor ($\nu_e$, ${\bar \nu}_{\mu}$ and $\nu_{\mu}$) of neutrino is 
emitted with $\sim 5\%$ of the proton energy. Using the standard internal shock 
model, as in \cite{Waxman+97grbnu}, the neutrino spectrum has a spectral break at 
an  energy $E_{\nu,b} \sim 10^{15}$ eV, where the neutrino production efficiency 
is high. This break is related via the $\Delta$-resonance condition 
and the bulk Lorentz factor to the photon spectral break energy $E_{\gamma,b} 
\sim 1$ MeV of the Band spectrum. For a generic photon spectrum with slopes 
$dN(E_\gamma)/dE_\gamma \propto E_\gamma^{-1,-2}$ below/above $E_{\gamma,b}$,
and protons with a spectrum $N(E_p)\propto E_p^{-2}$, the neutrino spectrum 
coincidentally has slopes $dN(E_\nu)/dE_\nu\propto E_\nu^{-1,-2}$ below/above
$E_{\nu,b}\sim$ PeV; for other photon and proton spectral slopes the neutrino 
slopes are different, but qualitatively similar. The fluxes of all three neutrino 
flavors ($\nu_e$, $\nu_{\mu}$ and $\nu_{\tau}$) are expected to be equal
after oscillation in vacuum over astrophysical distances.  The diffuse muon 
neutrino flux from GRB internal shocks  in this scenario, using an average
GRB photon luminosity, spectrum and bulk Lorentz factor as well as standard 
GRB occurrence statistics, is expected to be comparable or somewhat below 
the so-called Waxman-Bahcall (WB) diffuse neutrino flux for optically (neutrino) 
thin sources implied by the observed cosmic ray flux \cite{Waxman+00nuag}. 

This GRB ultra-high energy neutrinos (UHENU) scenario, based on internal
shocks, was used by \cite{Guetta+04grbnu} to predict a diffuse neutrino flux
which scales with the MeV photon fluxes of GRBs detected by a spacecraft.
This can be done having observed the photon spectra, and using these for
predicting a  neutrino flux (same duration as the photon burst) by assuming
neutrino production via the $\Delta$-resonance, a proton spectrum $\propto E_p^{-2}$
and a relativistic proton energy scaling with the relativistic electron energy 
by a factor $f_p=1/f_e=({\cal E}_p/{\cal E}_e) \sim 10$, where the electron energy 
is assumed, due to the fast cooling, to be equal to the observed MeV photon energy. 
The neutrino spectral shape has a break which scales with observed photon spectral 
break, and thus using the actual MeV fluxes and spectra of GRBs measured by a 
spacecraft they predict a cumulative neutrino flux over the period of observations, 
say a year. 

\begin{figure}[htb]
\begin{minipage}{0.65\textwidth}
\includegraphics[width=1.0\textwidth,height=2.5in,angle=0.0]{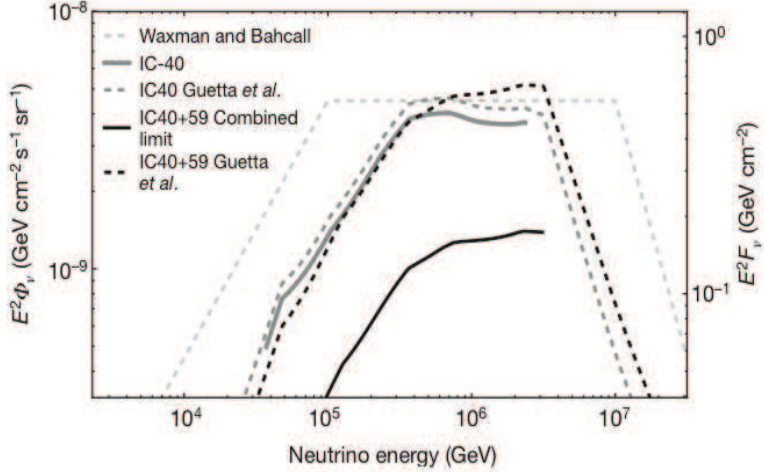}
\end{minipage}
\hspace{5mm}
\begin{minipage}[t]{0.32\textwidth}
\vspace*{-1.35in}
\caption{IceCube upper limits \cite{Abbasi+12grbnu-nat} for the 40-string array 
and the 40+59 string array on 190 GRBs localized by \swift, compared to an 
internal shock model \cite{Guetta+04grbnu} scaled to the photon luminosity 
(right y-axis).  The data is a factor 3.7 below this model.  The left y-axis is the 
diffuse flux calculated for 677 bursts/year, the WB limit is in light dashed lines.}
\end{minipage}
\label{fig-ic3}
\end{figure}
Recently, the IceCube group has been analyzing the TeV-PeV UHENU data accumulated 
by the 40-string array and subsequently the 59-string array \cite{Abbasi+11-ic40diff,
Abbasi+12grbnu-nat}, and comparing it to the GRB neutrino model flux expectations 
based on \cite{Guetta+04grbnu}. In the combined 40 and 59 string data (IC-40+59)
they used 190 electromagnetically detected bursts (Fig. 6), 
with their specific photon spectral slopes and breaks to predict, modulo the
factor $f_p=10$, the predicted neutrino spectra, adding them all up. The predicted
neutrino cumulative spectrum is shown in Fig. 6 for the 40 string array as the
gray dashed (model) and gray solid (data) lines; and for the 40+59 string array
as the dark dashed (model) and dark solid (data) lines; the original generic
Waxman-Bahcall spectrum is shown by the very light gray dashed lines. It is seen
that the IC-40+59 observed data fall a factor $0.27$ below the predicted model
neutrino flux. This means that this internal shock model over-predicts the data
by a factor 3.7, assuming a proton to electron energy ratio of 10, interactions via 
the $\Delta$-resonance and buck Lorentz factors in the usual range $\Gamma\sim 300-600$.
Using similar assumptions but not using the spectral information, if these GRB
supply the GZK cosmic ray flux, the implied neutrino flux is $2-3\sigma$ above
the level allowed by the data.  The conclusion is that either $f_p=({\cal E}_p/
{\cal E}_e)$ is  significantly below 10, or the production efficiency of neutrinos 
is lower than was assumed - or the specific model as  used can be largely ruled out.
This is a major landmark, being the first time that a specific extragalactic 
astrophysical source model is being tested through neutrino observations, at this 
$\sim 95\%$ level of significance. IceCube is doing exciting astrophysics, and this 
is a major step towards testing a candidate astrophysical source of GZK cosmic rays. 

\begin{figure}[htb]
\begin{minipage}{0.65\textwidth}\includegraphics[width=1.0\textwidth,height=3.0in,angle=0.0]{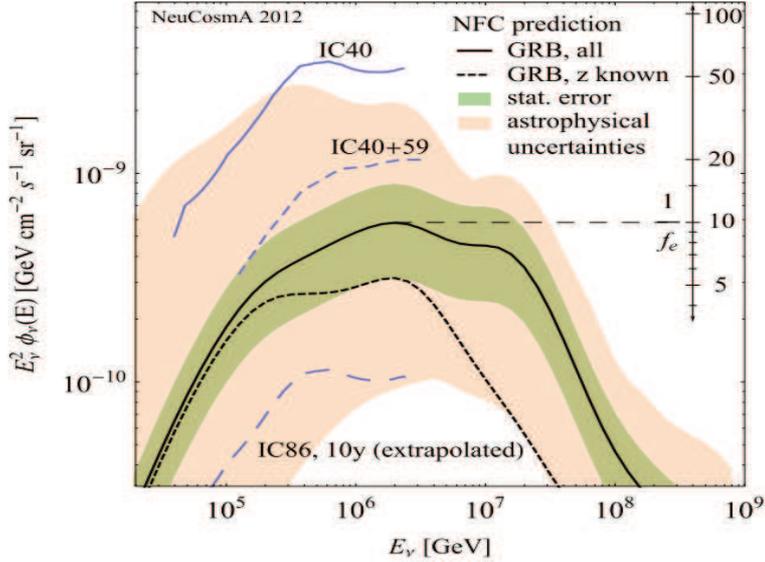}
\end{minipage}
\hspace{5mm}
\begin{minipage}[t]{0.3\textwidth}
\vspace*{-1.5in}
\caption{Predictions of the same internal shock model neutrino flux without
several of the previous approximations (solid and dashed dark lines), compared
to IC-40 and IC-40+59 data. The statistical error is given by the darker shaded 
region, the astrophysical uncertainties is given by the lighter shaded area 
\cite{Hummer+11nu-ic3}.  }
\end{minipage}
\label{fig-hummer}
\end{figure}
There are two additional points in this connection. One is that in using the 
standard internal shock model to predict the expected neutrino flux 
\cite{Abbasi+11-ic40diff,Abbasi+12grbnu-nat} made a number of simplifications. 
As discussed in \cite{Li11-nu-ic3}, the neutrino production efficiency at all 
energies was assumed to be the same as that evaluated at the break energy, but 
integrating the efficiency over the actual photon spectrum below and above the 
break yields a total neutrino flux a factor five lower than that used in the 
comparison with the data. 
Also, as discussed in \cite{Hummer+11nu-ic3,He+12grbnu}, including
production not just via $\Delta$-resonance but also via multi-pion and Kaon 
production, the spectrum is harder and this results in lower predicted fluxes
at the lower energies sampled.  Allowing for the  statistical uncertainty,
the $\pm 1\sigma$ limits of the predictions are calculated to be still below
the current IC-40+59 data, with $f_p=1/f_e=10$ (see Fig. 7). 
Considering the range of variability of the various astrophysical parameters, 
it appears that at least several years of observations with the full array may
be needed to reach conclusive results about the standard internal shock model. 
The other point is that internal shocks, in their standard form used above, have 
been known for a while to have some problems as far as efficiency and spectrum 
of the prompt $\gamma$-ray emission (see discussion in Sections 4 and 5). 
For this reason, alternative  models of MeV photon production have been 
considered, and the neutrino fluxes and spectra from such models are still
in the process of evaluation (e.g. \cite{Asano+12grbhad}).

\ack
We acknowledge partial support from NASA NNX08AL40G, NSF PHY0757155 and
Grant-in-Aid for Scientific Research No.22740117 from the Ministry of Education,
Culture, Sports, Science and Technology (MEXT) of Japan.


\section*{References}
\providecommand{\newblock}{}


\end{document}